\documentclass[proceedings, preprint]{rmaa}

\usepackage{paralist}

\usepackage{psfrag,color}

\usepackage{multirow}

\SetYear{2022}
\SetConfTitle{Prospects for low-frequency radio astronomy in S. A.}

\title{Prospects for detecting fast transients with the radio telescopes of the Argentine Institute of Radio Astronomy} 

\author{
  S. B. Araujo Furlan,\altaffilmark{1,2} 
  E. Zubieta,\altaffilmark{3,4}
  G. Gancio,\altaffilmark{3}
  G.E. Romero,\altaffilmark{3}
  S. del Palacio,\altaffilmark{3,5}
  F. García,\altaffilmark{3}
C. O. Lousto,\altaffilmark{6}
 and J. A. Combi\altaffilmark{3,4,7}}

\altaffiltext{1}{Instituto de Astronom\'\i{}a Teórica y Experimental, CONICET-UNC, Laprida 854, X5000BGR – Córdoba, Argentina 
(saraujo@iar.unlp.edu.ar).}

\altaffiltext{2}{Facultad de Matemática, Astronomía, Física y Computación, UNC. Av. Medina Allende s/n , Ciudad Universitaria, CP:X5000HUA - Córdoba, Argentina. }

\altaffiltext{3}{Instituto Argentino de Radioastronom\'\i{}a, CONICET–
CICPBA–UNLP. Cno. Gral. Manuel Belgrano
km 40, Pereyra, Buenos Aires, Argentina.}

\altaffiltext{4}{Facultad de Ciencias Astronómicas y Geofísicas, Universidad Nacional de La Plata. Paseo del Bosque s/n, FWA, B1900, Argentina.}

\altaffiltext{5}{Department of Space, Earth and Environment, Chalmers University of Technology. SE-412 96 Gothenburg, Sweden.}

\altaffiltext{6}{Center for Computational Relativity and Gravitation (CCRG), School of Mathematical Sciences, Rochester Institute of Technology, 85 Lomb Memorial Drive, Rochester, New York 14623, USA.}

\altaffiltext{7}{Departmento de Física, Universidad de Jaén. Campus Las Lagunillas s/n, 23071 Jaén, España.}

\shortauthor{Araujo Furlan et al.}
\shorttitle{Prospects for detection of radio transients}

\listofauthors{S. B. Araujo Furlan, E. Zubieta, G. Gancio, G. E. Romero, S. del Palacio, F. Garcia, C. O. Lousto, J. A. Combi}
\indexauthor{Araujo Furlan, S. B.}
\indexauthor{Zubieta, E.}
\indexauthor{Gancio, G.}
\indexauthor{Romero, G. E.}
\indexauthor{del Palacio, S.}
\indexauthor{García, F.}
\indexauthor{Lousto, C. O.}
\indexauthor{Combi, J. A.}

\abstract{

 Currently, 6 out of 30 known magnetars had pulsed radio emission detected. In this work, we evaluated the possibility of detecting radio transient events from magnetars with the telescopes of the Instituto Argentino de Radioastronomía (IAR). To this aim, we made daily observations of the magnetar XTE~J1810$-$197 from 02-Sep-22 to 30-Nov-22.
 We analysed the observations by applying ephemeris folding and single pulse searches. We fitted a timing model to our observations and were able to detect the magnetar on 6 of the 36 observing sessions with signal-to-noise ratios at the limit of detectability, $3.3\leq \mathrm{S/N} \leq4.1$. We searched for individual pulses in one of these 6 days and found 7 individual pulses with $8.5\leq \mathrm{S/N} \leq18.8$. The dispersion measure changed slightly between pulses within a range of $178 \leq \textrm{DM} \,[\mathrm{pc\, cm^{-3}}] \leq 182$. The pulse with $\mathrm{S/N}=18.8$ has an associated $\textrm{DM}$ of $180\,\mathrm{pc\, cm^{-3}}$. We confirmed that we can detect pulsed radio emission in the band of $1400-1456\, \mathrm{MHz}$ from magnetars with a time resolution of $146\,\mu s$, being able to detect both integrated pulse profiles and individual pulses. 

}

\resumen{

 Actualmente, s\'olo 6 de los 30 magnetares conocidos han tenido emisión en radio reportada.
 En este trabajo, evaluamos la posibilidad de detectar eventos transitorios en radio provenientes de magnetares, usando los telescopios del Instituto Argentino de Radio Astronomía (IAR). 
 Para ello, realizamos observaciones diarias del magnetar XTE~J1810$-$197 desde el 02-Sep-22 al 30-Nov-22. Analizamos las observaciones empleando integraciones en fase y búsqueda de pulsos individuales. 
 Ajustamos un modelo de \textit{timing} actualizado para nuestras observaciones y detectamos al magnetar en 6 de 36 días observados con una señal a ruido de $3.3\leq \mathrm{S/N} \leq 4.1$. Buscamos pulsos individuales en uno de estos 6 días y encontramos 7 pulsos con $8.5\leq \mathrm{S/N} \leq18.8$.  
La medida de dispersi\'on vari\'o ligeramente entre los pulsos dentro del rango $178 \leq \textrm{DM} \,[\mathrm{pc\, cm^{-3}}] \leq 182$. 
El pulso con $\mathrm{S/N}=18.8$ tiene una $\textrm{DM}$ asociada de $180\,\mathrm{pc\, cm^{-3}}$.
Demostramos que podemos detectar emisión pulsada en radio ($1400-1456\, \mathrm{MHz}$) de magnetares con una resolución temporal de $146\,\mu s$, observando perfiles de pulsos integrados y pulsos individuales. 

}

\addkeyword{Methods: observational}
\addkeyword{Methods: data analysis}
\addkeyword{Stars: magnetars}
\addkeyword{Stars: neutron}
\addkeyword{Radio continuum: general}

\begin{document}
\maketitle

\section{Introduction} 
\label{sec:intro}

Magnetars are a particular class of young, slowly rotating neutron stars ($P\sim 1$--$12\,\mathrm{s}$) with extremely strong surface magnetic fields ($B\sim10^{13}$--$10^{15}\,\mathrm{G}$). They exhibit a rich transient phenomenology, showing giant flares, short bursts, and outbursts, detected mainly at X-rays \citep{2017ARA&A..55..261K}. The energy for the observed X-ray and $\gamma$-ray emission is provided by the decaying magnetic fields \citep{1992ApJ...392L...9D}. Only 6 out of the 30 known magnetars had radio emission detected so far \citep{2014ApJS..212....6O}\footnote{\url{https://www.physics.mcgill.ca/~pulsar/magnetar/main.html}}. Five of them presented detectable transient radio pulsations, always associated with X-ray outbursts. The remaining one, SGR 1935+2154, showed Fast Radio Burst (FRB)-like bursts  \citep{2020Natur.587...54C,2020Natur.587...59B}. 

Studying the pulsed radio emission of magnetars is a tool to probe their spectral and temporal phenomenology. This emission has a switch on-off behaviour, going through quiescent states. It also shows great pulse-to-pulse variability, in both single and averaged pulses (\citealp[for recent reviews]{2017ARA&A..55..261K, 2021ASSL..461...97E}). Additionally, magnetars have been long suspected to be the source of FRBs. This hypothesis has been strongly supported by the recent detection of a FRB-like episode from SGR 1935+2154. Magnetars have been detected over a wide range of frequencies. For example, the magnetar XTE~J1810$-$197 has been detected in frequencies as low as 300 MHz \citep{2022ApJ...931...67M} and as high as 353 GHz \citep{2022ApJ...925L..17T}. 

In this context, we aim to assess the observational capabilities offered by the radio telescopes at the Argentine Institute of Radio Astronomy (IAR, after the acronym in Spanish), with a special focus on FRB-like phenomena. Previous efforts on the detection of magnetars were reported in \citet{2018ATel12323....1D}. There we reported the result of observations of the magnetar XTE~J1810$-$197 following its last outburst in late 2018.   

We present an analysis of the prospects for detecting transient radio emissions from magnetars and compact objects with the recently updated radio telescopes Carlos M. Varsavsky (A1) and Esteban Bajaja (A2) at IAR \citep{2020A&A...633A..84G}. These instruments can observe sources within a declination range of $-90\arcdeg<\delta<-10\arcdeg$, with a maximum time on-source of $\sim 3\mathrm{h}\, 40\mathrm{m}$.  

For this preliminary investigation, we observed XTE~J1810$-$197, since this source had previously been detected with both A1 and A2 \citep{2018ATel12323....1D}. XTE~J1810$-$197 was the first magnetar found to emit in radio \citep{2006Natur.442..892C}. It had two outbursts, the most recent one in late 2018 \citep{2018ATel12284....1L}. Different studies were performed on the time variability of the radio average profile, single pulses, flux density, and spectral index in the years following this outburst \citep{2019ApJ...874L..14D, 2019ApJ...882L...9M, 2019MNRAS.488.5251L, 2021PASJ...73.1563E, 2022MNRAS.510.1996C, 2022ApJ...925L..17T, 2022ApJ...931...67M}. The most recent study incorporates observations until April 2021 \citep{2022ApJ...931...67M}. An interesting result was the detection of Giant Pulses (GP)-like emission \citep{2019ApJ...882L...9M} and GP \citep{2022MNRAS.510.1996C} from this magnetar. 

Here we present a summary of the main results of the observations of XTE~J1810$-$197 made in the second half of 2022 from IAR.

\begin{table}[!t]
    \centering
      \setlength{\tabnotewidth}{0.45\columnwidth}
 \tablecols{2}
  \setlength{\tabcolsep}{0.8\tabcolsep}
  \caption{Timing Solution Parameters \tabnotemark{}} \label{tab:para}
    \begin{tabular}{ll}
    \toprule
       Parameter & Value \\ 
       \midrule
        Date range [MJD] & 59850.9--
        59895.8 \\
        DM [$\mathrm{pc\,cm^{-3}}$] & 178 \\ 
        \midrule
        Epoch of frequency [MJD] & 59849.9 \\ 
        F0 [$\mathrm{Hz}$] & 0.180420036(6) \\ 
        F1 [$\mathrm{Hz/s}$] & $-2.61(3)\times 10^{-13}$ \\
        \bottomrule
    \tabnotetext{}{}
    \end{tabular}
\end{table}

\begin{figure}[!t]
   \centering
  \includegraphics[width=0.85\columnwidth, trim=0.5cm 0.6cm 0.4cm 0.0cm, clip]{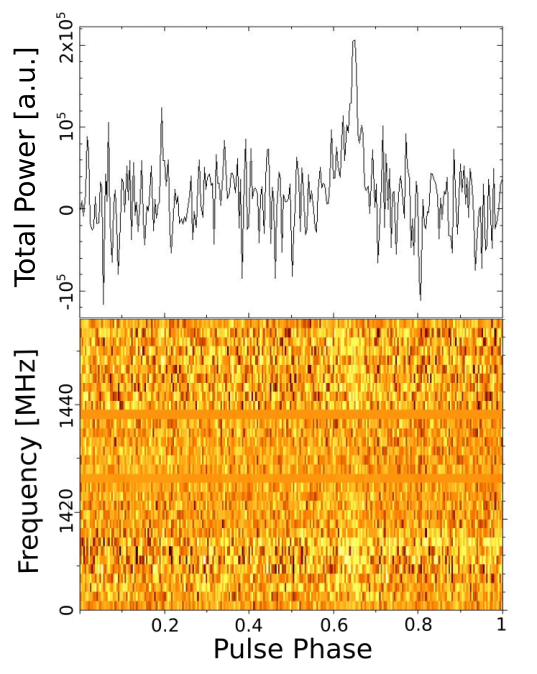}
  \caption{Integrated pulsed profile obtained for 29-Sep-22. In the upper panel, we show the integrated pulse. The horizontal axis represents one period of the magnetar expressed as pulse phase, and the vertical axis is the total power on arbitrary units. In the bottom panel we show the waterfaller plot (pulse phase versus frequency).}
  \label{fig:ip}
\end{figure}

\section{Observations}

The IAR radio observatory 
is located in the Pereyra Iraola Park in Argentina. 
It has two single dish radio telescopes of 30-metre diameter each.

The acquisition module for each antenna has been updated during the past couple of years \citep{2020A&A...633A..84G}. Nowadays, we routinely perform observations with ETTUS boards on the receivers. Each antenna has two of these boards, with a bandwidth of 56~MHz for each polarisation. The boards of A1 are configured as consecutive bands; this configuration gives a resulting bandwidth of 112~MHz with a total of 128 channels for a single polarisation. In the case of A2, the boards are configured to add both polarisation signals, measuring total power; 
this setting has a total bandwidth of 56~MHz. Both instruments have the same maximum observing frequency, $1456\, \mathrm{MHz}$.

We started a high-cadence monitoring campaign of the source on 02-Sep-22 with A1 and on 27-Sep-22 with A2. The observations were taken with a sampling time of $146\, \mu\mathrm{s}$, a frequency resolution of 0.875~MHz\footnote{During 27-Sep-22 to 02-Oct-22 the acquisition configuration was set to 32 channels with a resolution of 1.75-MHz.} and an average total time on source of $2.4~\mathrm{h}$. Observations with A1 were strongly affected by radio frequency interference (RFI). We thus focused the analysis on observations made with A2 as they were cleaner. The time span analysed goes from 27-Sep-22 until 30-Nov-22, with a total of 37 days observed, summing $\sim72~\mathrm{h}$ on source.

\section{Reduction and data analysis}
\label{sec:reduc}

We used two independent methods in searching for emission from the magnetar: one for obtaining an integrated pulse profile, and another one for searching for single pulses bright enough as to be individually detected. 

\subsection{Ephemeris folding}

The search for integrated pulse profiles was done with the PuMA pipeline\footnote{\url{https://github.com/PuMA-Coll/PuMA}}. The pipeline makes an RFI excision with the task \texttt{rfifind} from \texttt{PRESTO}\footnote{\url{https://github.com/scottransom/presto}} and then it folds the observation with \texttt{prepfold}. The signal-to-noise ratio (S/N) for the magnetar was very low to fit its period each day. Instead, we 
made an iterative process consisting of i) folding the observations with the most updated timing solution, ii) computing the time of arrival (TOA) for the observations with S/N $\gtrsim 3$, and iii) fitting the residuals with \texttt{Tempo2} \citep{2006MNRAS.369..655H} to improve the timing model. In step i), we used as a seed ephemeris  
the one reported in \citet{2022MNRAS.510.1996C}. 
In step ii), we calculated the S/N using \texttt{PyPulse} \citep{2017ascl.soft06011L}.
We iterated this process until the timing solution converged.

\subsection{Search of single pulses}

We followed the process described in \citet{2019ApJ...874L..14D,2019ApJ...882L...9M,2020ApJ...896L..30E, 2022ApJ...931...67M, 2022MNRAS.515.3577H} to detect single pulses and radio transients. First, we corrected the data for the dispersion caused by the interstellar medium. \texttt{PRESTO}'s \texttt{prepsubband} task corrects for dispersion a filterbank file and creates time series for each dispersion measure (DM) value in a previously specified range. The radio pulse emission from the magnetar should have the highest S/N for the dedispersed time series at the magnetar's DM. Each time series is obtained as a .dat and a .inf file. We used the mask previously created by the PuMA pipeline for RFI excision. We made 400 time series, with a DM step of $1\, \mathrm{pc\, m^{-3}}$, starting at a DM of $100\, \mathrm{pc\, m^{-3}}$. We searched within a broader range of DMs to search also for other transient events, such as FRBs. 

We then searched for single pulses within each time series. We employed \texttt{single\_pulse\_search.py} from \texttt{PRESTO}, with an S/N threshold of 8. This script returns a list of candidates for each time series. The output list can contain both astrophysical signals and RFIs. It also makes a diagnostic plot of the DM value, the S/N and the time for each candidate. We visually inspected this plot. To discriminate RFIs we employed \texttt{SpS}, a machine learning classifier designed to identify astrophysical signals from RFI \citep{2018MNRAS.480.3457M}, optimised for the discrimination of FRBs. The final results of \texttt{SpS} are the single pulses most likely to be of astrophysical origin. We searched with \texttt{SpS}' \texttt{no\_filter} option and without it. With this option, it does not filter the pulses reported. We cross-checked the diagnostic plot of \texttt{single\_pulse\_search.py} and the results of \texttt{SpS}.

\begin{table}[!t]
      \centering
    \setlength{\tabnotewidth}{0.6\columnwidth}
 \tablecols{4}

  \caption{Single Pulses detected on 29-Sep-22 \tabnotemark{}} 
  \label{tab:sps}
    \begin{tabular}{lccc}
    \toprule

        MJD & Time from  & DM & S/N \\ 
        &the start [s] & [$\mathrm{pc\, cm^{-3}}$] \\
        \midrule
        59851.895922569 & 1649.71 & 180 & 18.8\\ 
        59851.904454977 & 2386.91 & 178 & 9.8\\ 
        59851.90567338 & 2492.18 & 180 & 8.5\\
        59851.919273843 & 3667.26 & 178 & 8.7\\
        59851.926907292 & 4326.79 & 181 & 12.9 \\ 
        59851.94499838 & 5889.86 & 180 & 14.6\\ 
        59851.947563773 & 6111.51 & 182 & 9.0\\ 

        \bottomrule

         \tabnotetext{}{}
    \end{tabular}
\end{table}

\section{Results}

\begin{figure*}[!t]
\centering
 \includegraphics[width=\columnwidth,height=\columnwidth]{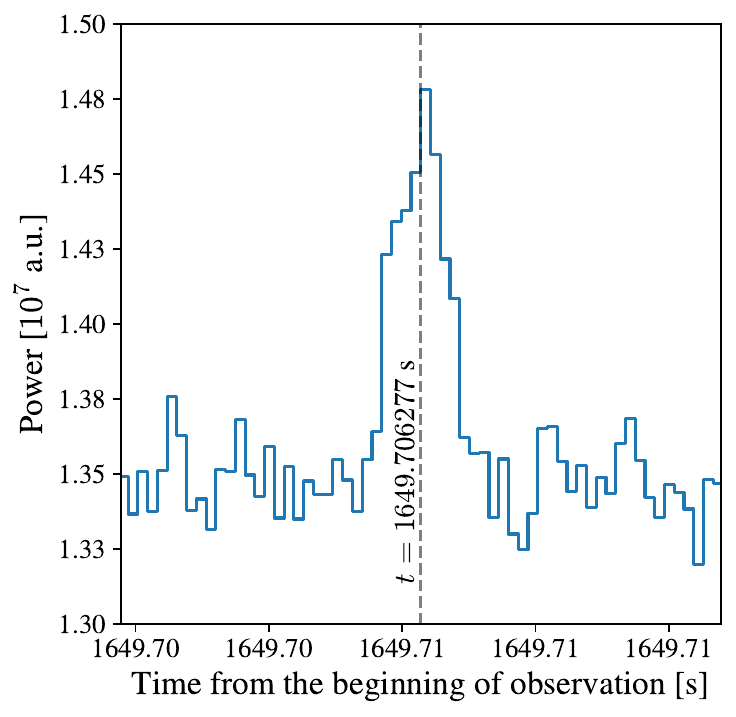}
  \includegraphics[width=\columnwidth,height=\columnwidth]{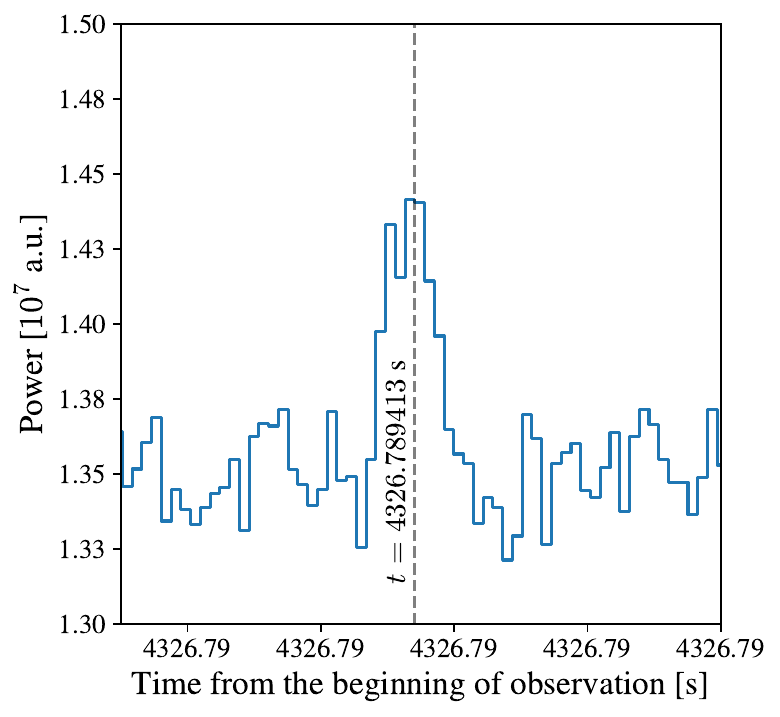}
  \caption{ Plot of two single pulses detected on 29-Sep-22 for the time series dedispersed at $\textrm{DM}=180\,\mathrm{pc\, m^{-3}}$. We plot time vs power in arbitrary units. The vertical line indicates the time of the pulse in seconds. }
  \label{fig:sp}
\end{figure*}

\subsection{Ephemeris folding}

We did the folding as described in \S~\ref{sec:reduc}. 
With the fitted timing model, we detected radio emission in 6 out of 37 days with S/N of 
$3.3\leq \mathrm{S/N} \leq4.1$. The parameter values of the model are given in Table~\ref{tab:para}. We detected pulsating periodic emissions on 27 and 29 of September, 18 and 21 of October and on the 03 and 06 of November. In Figure~\ref{fig:ip} we show the profile for 29-Sep-22, the day with $\mathrm{S/N}=4.1$. We can see the pulse centred at $\sim0.65$ of the pulse phase on the dynamic spectrum as in the integrated pulse profile. In the phase vs frequency plot, the pulse appears brighter on the outer parts of the bandwidth observed.

\subsection{Single pulse search} 

Here, we present the search for single pulses on the data taken on 29-Sep-22 with A2. The analysis of the remaining days will be reported in a future work. 

The diagnosis plot obtained with \texttt{single\_pulse\_search.py} showed 7 candidates centred around the magnetar's DM. When we employed \texttt{SpS} with the parameter \texttt{no\_filter}, it selected 6 of the pulses. Without that option, it recognised only 2. In Table~\ref{tab:sps} we present the MJD, the time of the pulse from the start of the observation, the S/N, and the DM at which \texttt{SpS} obtained the best S/N. The pulse at $t=2492.18\,\textrm{s}$ was the one that \texttt{SpS} did not recognise in either search. The remaining were recognised with \texttt{SpS}' \texttt{no\_filter} parameter. The only reported pulses without the \texttt{no\_filter}
option were the ones at $t=1649.71 \,\textrm{s}$ and $t=4326.79 \,\textrm{s}$.

We visually inspected the time series for $\textrm{DM}=180\,\mathrm{pc\, m^{-3}}$, as it is the assigned DM value for the highest S/N pulse. In Figure~\ref{fig:sp} we show the pulses of $\textrm{S/N}=18.8$ and  $\textrm{S/N}=12.9$. This latter one had that S/N at an associated $\textrm{DM}=181\,\mathrm{pc\, m^{-3}}$. We can see on the right panel that the pulse is recognisable at another DM. The left panel shows the highest S/N pulse. The vertical axis is power in arbitrary units as we did not have a flux calibration. The pulse centred at $t=1649.71\,\textrm{s}$ extends for $\sim13$ time bins, corresponding to $(1.9\pm0.1 )\, \mathrm{ms}$, while the pulse at $t=4326.79 \,\textrm{s}$ extends for 11 time bins, that is, $(1.6\pm 0.1)\, \mathrm{ms}$.

\section{Discussion and Conclusions}

We explored the possibility of detecting fast radio transients with the radio telescopes of the Argentine Institute of Radio Astronomy. 

We detected integrated pulse profiles from the magnetar XTE~J1810$-$197 on 6 out of 37 days of observations with S/N close to the limit of detectability. The emission probably corresponds to the afterglow of the 2018 outburst.

As a successful detection of the integrated pulses profile greatly depends on the employed timing solution, improving our model will lead us to better pulse profiles and more detections. As shown in Table~\ref{tab:para}, we use a rather simple model. Increasing the complexity of the model and upgrading it to our observations will be necessary for improvement. 

We successfully detected 7 single pulses from XTE~J1810$-$197 on the observation made on 29-Sep-22, with a sampling time of $146\, \mu s$, and with $8<\mathrm{S/N}$ for each of the single pulses. The values of S/N for each pulse are shown in Table~\ref{tab:sps}. 

\texttt{SpS} changes the number of reported pulses if we employ the \texttt{no\_filter} parameter, as this tells it to not filter pulses. We noticed that if not used, we obtained only 2 of the 6 pulses reported when we use it. All the pulses were centred around the magnetar's DM within a range of $178 \leq \textrm{DM} \,[\mathrm{pc\, cm^{-3}}] \leq 182$. The DM shown in Table \ref{tab:sps} is the value for which \texttt{SpS} obtains the highest $\mathrm{S/N}$ for the pulses. The remaining pulses reported with \texttt{no\_filter} have $8.5\leq\textrm{S/N}\leq14.6$.  The lower limit of $\textrm{S/N}=8$ corresponds to the chosen threshold S/N for \texttt{single\_pulse\_search.py}. As the pulses reported differ from \texttt{SpS} and \texttt{single\_pulse\_search.py}, we concluded that a visual inspection of the diagnosis plot and the time series is necessary for a correct interpretation of the results. 

As our observations have an important presence of RFI, especially A1's, we want to study the effect of employing other RFI excision processes for our search for single pulses. In \citet{2022MNRAS.509.5790L} they concluded that using \texttt{rficlean} plus \texttt{rfifind} increased greatly the $\textrm{S/N}$ of A1's pulse profiles, while for A2 the improvement was not that significant. We aim to study the benefits of employing both excisions for the search of individual pulses. 

Previous studies reported single pulse radio emission up to MJD 59300 \citep{2022ApJ...931...67M}. We detected single pulses emission on MJD 59851.8. We suspect by the value of the obtained $\textrm{S/N}$ for the pulses, that they may be Giant Pulses. A direct estimate of the peak flux density of the brightest pulse, employing the radiometer equation 
for a temporally resolved pulse \citep{2014MNRAS.445.3221M}: 
\begin{equation}
    S^{SP}_{\rm peak} = ({\rm S/N})_{\rm peak} \frac{2k_{B}T_{\rm sys}}{A_{e}(z)\sqrt{n_{\rm p}W\Delta\nu}}
\end{equation}
yields  $S\sim (41\pm4)\,\textrm{Jy}$. In this expression $T_{\rm{sys}}$ is the system temperature, $A_{e}(z)$ is the effective collecting area as a function of zenith angle $(z)$, $\Delta\nu$ is the observed bandwidth, $n_{\rm p}$ is the number of polarizations (two for A2) and $({\rm S/N})_{\rm peak}$ is the peak S/N of the pulse, corresponding to a smoothing optimum for its observed width of $W$. We did not smooth the time series; instead we used the reported $({\rm S/N})_{\rm peak}$ obtained with \texttt{single\_pulse\_search.py}, with the apparent width of the pulse and the aperture efficiency taken from \citet{2020A&A...633A..84G}.  

It seems clear that the pulses are of significant intensity. We are currently searching pulses for the remaining observed days of this campaign. The monitoring of magnetar XTE~J1810$-$197 is ongoing since September 2022.

We have demonstrated in this study that we are able to detect transient radio events with IAR's radio telescopes with a sampling time of $146\, \mu\textrm{s}$ at frequencies of $(1400-1456\,\textrm{MHz})$. 

\vspace{0.3cm}

{\bf Acknowledgements:} Araujo Furlan S.B. thanks Marcus E. Lower for the discussion regarding the source we observed in this study. We thank the staff of IAR for help with the instruments during the observing campaign.


\begin{thebibliography}

\bibitem[Bochenek et al.(2020)]{2020Natur.587...59B} Bochenek, C.~D., Ravi, V., Belov, K.~V., et al.\ 2020, \nat, 587, 59


\bibitem[Caleb et al.(2022)]{2022MNRAS.510.1996C} Caleb, M., Rajwade, K., Desvignes, G., et al.\ 2022, \mnras, 510, 1996

\bibitem[Camilo et al.(2006)]{2006Natur.442..892C} Camilo, F., Ransom, S.~M., Halpern, J.~P., et al.\ 2006, \nat, 442, 892

\bibitem[CHIME/FRB Collaboration et al.(2020)]{2020Natur.587...54C} CHIME/FRB Collaboration, Andersen, B.~C., Bandura, K.~M., et al.\ 2020, \nat, 587, 54

\bibitem[Dai et al.(2019)]{2019ApJ...874L..14D} Dai, S., Lower, M.~E., Bailes, M., et al.\ 2019, \apjl, 874, L14 

\bibitem[Duncan \& Thompson(1992)]{1992ApJ...392L...9D} Duncan, R.~C. \& Thompson, C.\ 1992, \apjl, 392, L9

\bibitem[Eie et al.(2021)]{2021PASJ...73.1563E} Eie, S., Terasawa, T., Akahori, T., et al.\ 2021, \pasj, 73, 1563

\bibitem[Esposito et al.(2020)]{2020ApJ...896L..30E} Esposito, P., Rea, N., Borghese, A., et al.\ 2020, \apjl, 896, L30

\bibitem[Esposito et al.(2021)]{2021ASSL..461...97E} Esposito, P., Rea, N., \& Israel, G.~L.\ 2021, Timing Neutron Stars: Pulsations, Oscillations and Explosions, 461, 97


\bibitem[Gancio et al.(2020)]{2020A&A...633A..84G} Gancio, G., Lousto, C.~O., Combi, L., et al.\ 2020, \aap, 633, A84. doi:10.1051/0004-6361/201936525
\bibitem[Dai et al.(2019)]{2019ApJ...874L..14D} Dai, S., Lower, M.~E., Bailes, M., et al.\ 2019, \apjl, 874, L14


\bibitem[Del Palacio et al.(2018)]{2018ATel12323....1D} Del Palacio, S., Garcia, F., Combi, L., et al.\ 2018, The Astronomer's Telegram, 12323
\bibitem[Hewitt et al.(2022)]{2022MNRAS.515.3577H} Hewitt, D.~M., Snelders, M.~P., Hessels, J.~W.~T., et al.\ 2022, \mnras, 515, 3577


\bibitem[Hobbs et al.(2006)]{2006MNRAS.369..655H} Hobbs, G.~B., Edwards, R.~T., \& Manchester, R.~N.\ 2006, \mnras, 369, 655

\bibitem[Kaspi \& Beloborodov(2017)]{2017ARA&A..55..261K} Kaspi, V.~M. \& Beloborodov, A.~M.\ 2017, \araa, 55, 261

\bibitem[Lam(2017)]{2017ascl.soft06011L} Lam, M.~T.\ 2017, Astrophysics Source Code Library


\bibitem[Levin et al.(2019)]{2019MNRAS.488.5251L} Levin, L., Lyne, A.~G., Desvignes, G., et al.\ 2019, \mnras, 488, 5251


\bibitem[Lousto et al.(2022)]{2022MNRAS.509.5790L} Lousto, C.~O., Missel, R., Prajapati, H., et al.\ 2022, \mnras, 509, 5790

\bibitem[Lyne et al.(2018)]{2018ATel12284....1L} Lyne, A., Levin, L., Stappers, B., et al.\ 2018, The Astronomer's Telegram, 12284

\bibitem[Maan \& Aswathappa(2014)]{2014MNRAS.445.3221M} Maan, Y. \& Aswathappa, H.~A.\ 2014, \mnras, 445, 3221


\bibitem[Maan et al.(2019)]{2019ApJ...882L...9M} Maan, Y., Joshi, B.~C., Surnis, M.~P., et al.\ 2019, \apjl, 882, L9


\bibitem[Maan et al.(2022)]{2022ApJ...931...67M} Maan, Y., Surnis, M.~P., Chandra Joshi, B., et al.\ 2022, \apj, 931, 67

\bibitem[Manchester et al.(2005)]{2005AJ....129.1993M} Manchester, R.~N., Hobbs, G.~B., Teoh, A., et al.\ 2005, \aj, 129, 1993

\bibitem[Michilli et al.(2018)]{2018MNRAS.480.3457M} Michilli, D., Hessels, J.~W.~T., Lyon, R.~J., et al.\ 2018, \mnras, 480, 3457

\bibitem[Olausen \& Kaspi(2014)]{2014ApJS..212....6O} Olausen, S.~A. \& Kaspi, V.~M.\ 2014, \apjs, 212, 6

\adjustfinalcols

\bibitem[Torne et al.(2022)]{2022ApJ...925L..17T} Torne, P., Bell, G.~S., Bintley, D., et al.\ 2022, \apjl, 925, L17


 
\end{thebibliography}
\end{document}